\pdfoutput=1
\documentclass{PoS}

\usepackage[intlimits]{amsmath}
\usepackage{braket}
\usepackage{colortbl}
\usepackage{multirow}
\usepackage{tikz}

\title{Dipole subtraction with random polarisations}

\ShortTitle{Dipole subtraction with random polarisations}

\author{%
	\speaker{Daniel Goetz}, Christopher Schwan, Stefan Weinzierl\\%
	PRISMA Cluster of Excellence, Johannes Gutenberg University, Mainz\\
	\email{goetz@uni-mainz.de}\\
	\email{schwan@uni-mainz.de}\\
	\email{stefanw@thep.physik.uni-mainz.de}
}

\abstract{%
	In this talk, we discuss the speed-up of numerical calculations of jet observables by replacing the usual sum over all helicity amplitudes with an integral over a parametrisation for the parton polarisations called \emph{random polarisations}.
	Random polarisations are a linear combination of helicity eigenstates multiplied by a phase factor depending on a so-called \emph{helicity angle}.
	Instead of a summation over discrete helicities, random polarisations require an integration over the helicity angle. By combining this integral with the final-state phase space integral, we only have to evaluate one squared amplitude per phase space point instead of $2^n$ helicity amplitudes, where $n$ is the total number of particles in the process.

	While the technique itself has been known since 1998, so far there has been no way of using it with dipole subtraction, which is probably the most-used method for dealing with infrared divergences in NLO calculations.
	After giving detailed reasons for this statement, we propose a solution to this problem in terms of extending the existing subtraction method by a new term.
}

\FullConference{%
	11th International Symposium on Radiative Corrections (Applications of Quantum Field Theory to Phenomenology) \\
	22-27 September 2013\\
	Lumley Castle Hotel, Durham, UK
}

\newcommand{\erm}{\mathrm{e}}
\newcommand{\drm}{\mathrm{d}}
\newcommand{\irm}{\mathrm{i}}
\newcommand{\ijToIJ}{\widetilde{ij} \to i + j}

\begin{document}

\section{Introduction}

In this talk, we are dealing with the evaluation of jet cross sections for unpolarised scattering using numerical methods.
Our focus therein lies on three major aspects:
\begin{itemize}
\setlength\itemsep{0pt}\setlength\parskip{0pt}
	\item we want to develop a \emph{general} algorithm resulting in a fully automated code,
	\item we want to reach high particle multiplicities, e.g.\ $2 \to 6, 7, 8$ particles and, most importantly,
	\item we want to make the calculations as fast as possible.
\end{itemize}
A general jet cross section can be written as follows:
\begin{equation}
	\label{eq:jet-cross-section-general}
	\sigma
	= \int_n \drm \sigma
	\propto \int \drm \phi_{n-2} \sum_{\lambda_1, \ldots, \lambda_n} \lvert \mathcal{A}_{\lambda_1 \ldots \lambda_n} \rvert^2 \, F_J^{(n)}
\end{equation}
where $\drm \phi_{n-2}$ denotes the integration measure for the final state phase space integral, $F_J^{(n)}$ is the jet-defining function for $n$ particles, and the rest of the integrand is the helicity summed squared matrix element.

\section{Numerical polarisations}

The computationally expensive quantity in equation \eqref{eq:jet-cross-section-general} is the helicity summed squared amplitude. The ``classical'' method of computing this quantity is to write out the sum and compute each of the resulting \emph{helicity amplitudes} separately:
\begin{equation}
	\sum_{\lambda_1, \ldots, \lambda_n} \lvert \mathcal{A}_{\lambda_1 \ldots \lambda_n} \rvert^2
	= \lvert \mathcal{A}_{+++ \ldots +} \rvert^2 + \lvert \mathcal{A}_{-++ \ldots +} \rvert^2 + \lvert \mathcal{A}_{+-+ \ldots +} \rvert^2 + \cdots + \lvert \mathcal{A}_{--- \ldots -} \rvert^2 .
\end{equation}
In total, there are $2^n$ helicity amplitudes\footnote{In this talk, we restrict ourselves QCD and QED where all particles have two spin/helicity settings; if one considers e.g.\ electroweak processes with external on-shell (non-decaying) massive vector bosons, one has to split $n$ into the number $n_2$ of particles with two spin/helicity settings and the number $n_3$ of particles with three spin/helicity settings.
The total number of helicity amplitudes is then $2^{n_2} \cdot 3^{n_3}$.} that have to be computed per phase space point.
For high particle multiplicities, this clearly poses a problem if one wants a fast computation.
We would like to mention two methods that allow the speed-up of helicity amplitudes.
\begin{itemize}
\setlength\itemsep{0pt}\setlength\parskip{0pt}
	\item For parity conserving theories we can use the relation $\mathcal{A}_{\lambda_1 \lambda_2 \ldots \lambda_n} = - \mathcal{A}_{-\lambda_1, -\lambda_2, \ldots, -\lambda_n}^*$. Thus, it is sufficient to compute only half of all helicity amplitudes, i.e.\ we have only $2^{n - 1}$ amplitudes.
	\item Within recursive methods it is possible to store off-shell currents and reuse them in the computation of various helicity amplitudes.
\end{itemize}
While these two methods provide some speed-up, we want to go further: we can reduce the number of squared amplitudes per phase space point down to one by replacing the helicity sum with an integral over so-called \emph{helicity angles} $\theta$,
\begin{equation}
	\label{eq:helicity-sum-to-integration}
	\sum_{\lambda_1, \ldots, \lambda_n} \lvert \mathcal{A}_{\lambda_1 \ldots \lambda_n} \rvert^2
	= \int_{[0,1)^n} \drm^n \theta \, \lvert \mathcal{A}_{\theta_1 \ldots \theta_n} \rvert^2.
\end{equation}
To evaluate $\mathcal{A}_{\theta_1 \ldots \theta_n}$ we use the following parametrisation for the particle polarisation vectors \cite{Draggiotis:1998gr, Czakon:2009ss}:
\begin{equation}
	\label{eq:random-polarisations}
	\epsilon (p, \theta)
	= \erm^{2 \pi \irm \theta} \epsilon_+ (p) + \erm^{-2 \pi \irm \theta} \epsilon_- (p).
\end{equation}
\pagebreak
We call this parametrisation \emph{random polarisations} (RP).
The reason that this works is that the typical polarisation sum is preserved, which can be seen by rewriting a squared amplitude as follows:
\begin{equation}
	\label{eq:matrix-element-with-polarisation-sum}
	\sum_{\lambda_1, \ldots, \lambda_n} \lvert \mathcal{A}_{\lambda_1 \ldots \lambda_n} \rvert^2
	= \prod_{m=1}^n \left[ \sum_{\lambda_m} \epsilon_{\lambda_m}^{\mu_m} ( \epsilon_{\lambda_m}^{\nu_m} )^* \right] \times \mathcal{M}_{\mu_1, \ldots, \mu_n, \nu_1, \ldots, \nu_n}.
\end{equation}
In a squared amplitude, we have one polarisation sum for each external particle multiplied by the remainder $\mathcal{M}$ (which contains vertices and propagators).
If we use RPs, we first have to replace the eigenstates with the parametrisation from \eqref{eq:random-polarisations}.
For the product we thus obtain
\begin{equation}
\label{eq:random-polarisation-product}
\begin{split}
	\epsilon_\lambda^\mu (\epsilon_\lambda^\nu)^*
	\to
	\epsilon^\mu (\theta) \bigl( \epsilon^\nu (\theta) \bigr)^*
	&= \epsilon_+^\mu (\epsilon_+^\nu)^* + \epsilon_-^\mu (\epsilon_-^\nu)^* + \erm^{4 \pi \irm \theta} \epsilon_+^\mu (\epsilon_-^\nu)^* + \erm^{-4 \pi \irm \theta} \epsilon_-^\mu (\epsilon_+^\nu)^*
	\\
	&= \sum_\lambda \epsilon_\lambda^\mu (\epsilon_\lambda^\nu)^* + \erm^{4 \pi \irm \theta} \epsilon_+^\mu (\epsilon_-^\nu)^* + \erm^{-4 \pi \irm \theta} \epsilon_-^\mu (\epsilon_+^\nu)^*,
\end{split}
\end{equation}
which is the usual polarisation sum plus two additional terms that mix the different helicity eigenstates of the same particle, something which is unique to RP and very important in the course of this talk; hence, we will refer to these terms as \emph{helicity mixing terms}.

Furthermore, we have to replace the sum in equation \eqref{eq:matrix-element-with-polarisation-sum} with the integration indicated in \eqref{eq:helicity-sum-to-integration} which gives us
\begin{equation}
\label{eq:random-polarisation-product-integration}
	\int_0^1 \drm \theta \, \epsilon^\mu (\theta) \bigl( \epsilon^\nu (\theta) \bigr)^*
	= \sum_\lambda \epsilon_\lambda^\mu (\epsilon_\lambda^\nu)^* + \int_0^1 \drm \theta \, \left( \erm^{4 \pi \irm \theta} \epsilon_+^\mu (\epsilon_-^\nu)^* + \erm^{-4 \pi \irm \theta} \epsilon_-^\mu (\epsilon_+^\nu)^* \right)
	= \sum_\lambda \epsilon_\lambda^\mu (\epsilon_\lambda^\nu)^*
\end{equation}
where the helicity mixing terms vanish due to Cauchy's theorem and we are left with nothing but the usual polarisation sum.

Although the notation used in the above discussion suggests a restriction to boson polarisation vectors, we use the symbol $\epsilon$ in a generalised way. For fermion spinors, one can use parametrisations analogous to \eqref{eq:random-polarisations} simply by replacing $\epsilon$ with $u$, $\bar{u}$, $v$, or $\bar{v}$ and the Lorentz indices $\mu$, $\nu$ with spinor indices $\alpha$, $\beta$.

\begin{figure}
	\includegraphics[width=0.48\textwidth, page=1]{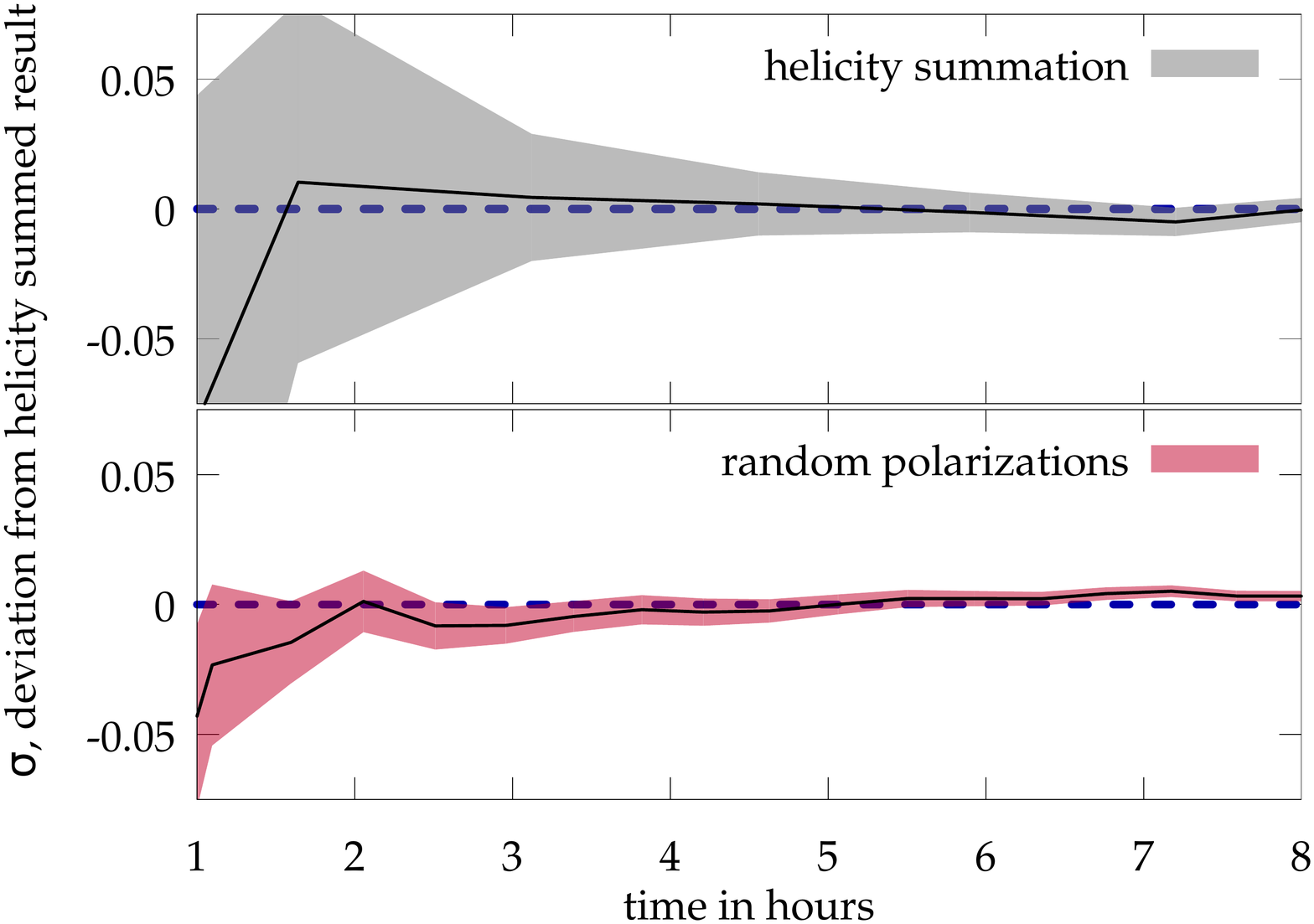} \hfill
	\includegraphics[width=0.48\textwidth, page=2]{pics/6jetsTalk}
	\caption{Comparison of helicity amplitudes and RP for $e^+ e^- \to 6$ jets at Born level.}
	\label{fig:6jets}
\end{figure}

What do we gain from using RP?
First of all, we reduce the number of squared amplitudes from $2^n$ down to one in comparison with the method of helicity amplitudes.
The additional $n$-dimensional integral can be combined with the already present integration over final-state phase space.
The question that arises immediately is how well this method with its additional integration dimensions performs in comparison with the traditional method helicity amplitudes.
Figure \ref{fig:6jets} shows a comparison between the two methods for the Born level cross section for $e^+ e^- \to \gamma^* \to 6$ Jets.
The left plot shows the integration results over a period of eight hours.
We do not show the absolute cross section but the deviation from the final helicity summed result.
The right hand plot shows the relative errors in percent plotted against the same time interval.
It is immediately obvious that RP are superior to the usual helicity summation.
Also note that the helicity summed results were generated using the two methods for speed-up mentioned earlier, so this is already an improved helicity summation that we are comparing against.

\section{Dipole subtraction}

At the end of the last section we saw that RP provide a great speed-up for leading order calculations.
However, collider energies such as they occur at the LHC require more precise calculations. This means that to provide accurate predictions for LHC physics, we need to go to next-to leading order whose contribution to the jet cross section reads\\[-0.5em]
\begin{equation}
	\sigma^{\text{NLO}} = \int_{n+1} \drm \sigma^{\text{R}} + \int_n \drm \sigma^{\text{V}}.
\end{equation}\\[-1em]
The fundamental difficulty with this contribtion is that both the real and the virtual contribution are separately divergent and only summing them after the integration cancels all the poles. In numerical calculations, we need a way to cancel the divergences locally, i.e.\ per phase space point. One ubiquitous way to do this is the subtraction method which introduces an additional subtraction term $\drm \sigma^{\text{A}}$ that acts as a local counter term for the infrared divergences:\\[-1em]
\begin{equation}
	\sigma^{\text{NLO}} = \int_{n+1} \left( \drm \sigma^{\text{R}} - \drm \sigma^{\text{A}} \right) + \int_n \left( \drm \sigma^{\text{V}} + \int_1 \drm \sigma^{\text{A}} \right).
\end{equation}
This $\drm \sigma^{\text{A}}$ must have the same pointwise singular behaviour as $\drm \sigma^{\text{R}}$, and it must be integrable over the unresolved one-parton phase space so that it can cancel the divergences of $\drm \sigma^{\text{V}}$ locally.

There are many ways to write down the subtraction term $\drm \sigma^{\text{A}}$.
In this talk we will focus on one of the most-used parametrisations, the Catani-Seymour dipole subtraction terms \cite{Catani:1996vz}.
Since we only need to subtract the divergent parts of the real emission amplitude, it is sufficient that $\drm \sigma^{\text{A}}$ mimics the poles.
The universality of infrared divergences provides us with what is called the external-leg insertion rule where the soft and collinear poles are approximated by removing one external leg from the amplitude and inserting it as a correlation parton.

In the dipole formalism, the subtraction terms take the final form\\[-1em]
\begin{equation}
\label{eq:dipole-subtraction-term}
	\drm \sigma^{\text{A}} \propto \drm \phi_{n-1} \left( \sum_{(i,j)} \sum_{k \neq i,j}
	\sum_{\{ \lambda \} \backslash \lambda_i, \lambda_j}
	\mathcal{D}_{ij,k} \, F_J^{(n)} + \text{initial state dipoles} \ldots \right)
\end{equation}
where the third sum runs over all particle helicities except those of particles $i$ and $j$.\footnote{Note that this also includes the summation over the helicities of a so-called emitter parton $\widetilde{ij}$ which will be explained in the following.}
Note that we restrict ourselves to final-state radiation in this talk.
The dipoles are given by
\begin{equation}
	\mathcal{D}_{ij,k} = \frac{1}{2 p_i p_j} \bra{\mathcal{A}_n (\widetilde{ij}, \widetilde{k})} \frac{\mathbf{T}_k \cdot \mathbf{T}_{ij}}{\mathbf{T}_{ij}^2} \mathbf{V}_{ij,k} \ket{\mathcal{A}_n (\widetilde{ij}, \widetilde{k})}.
\end{equation}
In essence, one removes partons $i$ and $j$ from the amplitude and replaces them with one so-called emitter parton $\widetilde{ij}$.
Its momentum is parametrised such that it is on-shell.
This only works if one other parton, the so-called spectator parton $k$, takes up some recoil:
\begin{equation}
	\widetilde{p}_{ij} = p_i + p_j - \frac{y_{ij,k}}{1 - y_{ij,k}} p_k, \quad
	\widetilde{p}_k = \frac{1}{1 - y_{ij,k}} p_k, \quad
	y_{ij,k} = \frac{p_i p_k}{p_i p_j + p_j p_k + p_k p_i}.
\end{equation}
The splitting of the emitter $\widetilde{ij}$ into $i$ and $j$ is then universally described by a spin-correlation matrix $\mathbf{V}_{ij,k}$.
This matrix is derived from the singular part of the squared, helicity summed splitting vertex.

Notice the sum over helicities that appears in $\drm \sigma^{\text{A}}$; the dipole formalism relies on the fact that one uses helicity summation.
This also means that for every dipole term in $\drm \sigma^{\text{A}}$, we have $2^{n-1}$ helicity dipoles.
In analogy to the Born level discussion in the last section, we would like to reduce computation time by reducing this number down to one --- the solution will be to use RP.

Let us investigate why it is not possible to trivially replace the sum over helicity eigenstates with an integration over random polarisations.
As mentioned before, the splitting matrix $\mathbf{V}_{ij,k}$ is derived from a \emph{helicity summed} splitting vertex; the actual form is not important here, so let us sketch it as follows:
{\small%
\begin{equation}
	\mathbf{V}_{ij,k} \sim
	\sum_{\lambda_i, \lambda_j}
	\left(
	\begin{tikzpicture}[scale=0.5, baseline=(thebaseline)]
		\node (thebaseline) at (0.0, -0.08) {};
		\draw (0.0, 0.0) -- (-1.0, 0.0) node[left=1pt] {$\widetilde{ij}$};
		\draw (0.0, 0.0) -- (0.7, 0.7) node[right] {$\epsilon_{\lambda_i}$};
		\draw (0.0, 0.0) -- (0.7,-0.7) node[right] {$\epsilon_{\lambda_j}$};
	\end{tikzpicture}
	\right)^*
	\left(
	\begin{tikzpicture}[scale=0.5, baseline=(thebaseline)]
		\node (thebaseline) at (0.0, -0.08) {};
		\draw (0.0, 0.0) -- (1.0, 0.0) node[right=1pt] {$\widetilde{ij}$};
		\draw (0.0, 0.0) -- (-0.7, 0.7) node[left] {$\epsilon_{\lambda_i}$};
		\draw (0.0, 0.0) -- (-0.7,-0.7) node[left] {$\epsilon_{\lambda_j}$};
	\end{tikzpicture}
	\right)
\end{equation}}%
In 2009, a generalisation of these splitting matrices to helicity eigenstates has been published \cite{Czakon:2009ss} (i.e.\ without the sum):
{\small%
\begin{equation}
	\mathbf{V}_{ij,k}^{\lambda_i \lambda_j} \sim
	\left(
	\begin{tikzpicture}[scale=0.5, baseline=(thebaseline)]
		\node (thebaseline) at (0.0, -0.08) {};
		\draw (0.0, 0.0) -- (-1.0, 0.0) node[left=1pt] {$\widetilde{ij}$};
		\draw (0.0, 0.0) -- (0.7, 0.7) node[right] {$\epsilon_{\lambda_i}$};
		\draw (0.0, 0.0) -- (0.7,-0.7) node[right] {$\epsilon_{\lambda_j}$};
	\end{tikzpicture}
	\right)^*
	\left(
	\begin{tikzpicture}[scale=0.5, baseline=(thebaseline)]
		\node (thebaseline) at (0.0, -0.08) {};
		\draw (0.0, 0.0) -- (1.0, 0.0) node[right=1pt] {$\widetilde{ij}$};
		\draw (0.0, 0.0) -- (-0.7, 0.7) node[left] {$\epsilon_{\lambda_i}$};
		\draw (0.0, 0.0) -- (-0.7,-0.7) node[left] {$\epsilon_{\lambda_j}$};
	\end{tikzpicture}
	\right),
	\qquad
	\begin{array}{ll}
		\lambda_i \in \{ +, - \}, \\
		\lambda_j \in \{ +, - \}
	\end{array}
\end{equation}}%
If we now recall that RP are linear combinations of helicity eigenstates, equation \eqref{eq:random-polarisations}, the question might arise why it is not possible to use the existing terms to construct splitting matrices for RP.
The answer is that the splitting matrices are based on the \emph{product} of polarisations which, in the above cases, includes $(\epsilon_+^\mu)^* \epsilon_+^\nu$ and $(\epsilon_-^\mu)^* \epsilon_-^\nu$.
RP however, lead to a more complicated product (see equation \eqref{eq:random-polarisation-product}) which includes new helicity mixing terms $(\epsilon_+^\mu)^* \epsilon_-^\nu$ and $(\epsilon_-^\mu)^* \epsilon_+^\nu$, one example term for particle $i$ could be represented as follows:
{\small%
\begin{equation}
	\mathbf{V}_{ij,k}^{\theta_i \theta_j} \sim
	\erm^{-4 \pi \irm \theta_i}
	\left(
	\begin{tikzpicture}[scale=0.5, baseline=(thebaseline)]
		\node (thebaseline) at (0.0, -0.08) {};
		\draw (0.0, 0.0) -- (-1.0, 0.0) node[left=1pt] {$\widetilde{ij}$};
		\draw (0.0, 0.0) -- (0.7, 0.7) node[right] {$\epsilon_+$};
		\draw (0.0, 0.0) -- (0.7,-0.7) node[right] {$\epsilon (\theta_j)$};
	\end{tikzpicture}
	\right)^*
	\left(
	\begin{tikzpicture}[scale=0.5, baseline=(thebaseline)]
		\node (thebaseline) at (0.0, -0.08) {};
		\draw (0.0, 0.0) -- (1.0, 0.0) node[right=1pt] {$\widetilde{ij}$};
		\draw (0.0, 0.0) -- (-0.7, 0.7) node[left] {$\epsilon_-$};
		\draw (0.0, 0.0) -- (-0.7,-0.7) node[left] {$\epsilon (\theta_j)$};
	\end{tikzpicture}
	\right)
	+ \ldots
\end{equation}}%
Since these terms also contribute poles to the overall local singularities, we cannot just ignore them.

\section{Random polarisations for real subtraction}

Now it is clear that something has to be done about these new terms. Let us remind ourselves once more about the product of two RP:
\begin{equation}
	\epsilon^\mu (\theta) \bigl( \epsilon^\nu (\theta) \bigr)^*
	= \sum_\lambda \epsilon_\lambda^\mu (\epsilon_\lambda^\nu)^* + \erm^{4 \pi \irm \theta} \epsilon_+^\mu (\epsilon_-^\nu)^* + \erm^{-4 \pi \irm \theta} \epsilon_-^\mu (\epsilon_+^\nu)^*,
\end{equation}
The first term is the usual polarisation sum of helicity eigenstates, which means that all singularities arising from this term are already taken care of by the existing subtraction terms $\drm \sigma^{\text{A}}$.
In fact, \emph{only} the helicity mixing terms are new and it is sufficient to construct new subtraction terms for these terms alone.
So, instead of deriving a completely new subtraction method we propose the following extension of the existing scheme:
\begin{equation}
	\sigma^{\text{NLO}}
	= \int_{n+1} \biggl( \drm \sigma^{\text{R}} - \left[ \drm \sigma^{\text{A}} + \drm \sigma^{\widetilde{\text{A}}} \right] \biggr) +
		\int_n \left( \drm \sigma^{\text{V}} + \int_1 \drm \sigma^{\text{A}} \right)
\end{equation}
The new term, $\drm \sigma^{\widetilde{\text{A}}}$ takes care of the helicity mixing terms.
Note that there is no need for a new integrated subtraction term.
Recall from equation \eqref{eq:random-polarisation-product-integration} that the helicity mixing terms vanish upon integration; this means that the accompanying integrated term has to vanish as well:
\begin{equation}
	\int_0^1 \drm \theta_i \int_0^1 \drm \theta_j \ \drm\sigma^{\widetilde{\text{A}}} = 0
\end{equation}
What is $\drm \sigma^{\widetilde{\text{A}}}$? Similar to the dipole formalism, this is given by a sum over new dipoles $\widetilde{\mathcal{D}}_{ij,k}$:
\begin{equation}
	\label{eq:random-dipole-subtraction-term}
	\drm \sigma^{\widetilde{\text{A}}}
	\ \propto \
	\drm \phi_{n - 1} \, \drm^{n+1} \theta \left( \sum_{(i,j)} \sum_{k \neq i,j} \widetilde{\mathcal{D}}_{ij,k} \, F_J^{(n)} + \text{ initial state dipoles} \ldots \right).
\end{equation}
In comparison to equation \eqref{eq:dipole-subtraction-term}, we replaced the helicity sum with an additional integration measure for the helicity angles.
The new dipoles also have a similar form as before,
\begin{equation}
	\label{eq:random-dipoles}
	\widetilde{\mathcal{D}}_{ij,k}
	= -4 \pi \alpha_s \mu^{2 \epsilon} \left( \mathcal{A}_n^* \right)^\xi \frac{\mathbf{T}_{ij} \cdot \mathbf{T}_k}{\mathbf{T}_{ij}^2} \left[ \widetilde{\mathbf{V}}_{ij,k} (\widetilde{p}_{ij}, p_i, p_j, p_k, \theta_i, \theta_j) \right]_{\xi \xi'} \mathcal{A}_n^{\xi'}.
\end{equation}
Notice the additional indices $\xi$, $\xi'$ of the amplitude and the new splitting matrices $\widetilde{\mathbf{V}}_{ij,k}$; here, we remove the polarisation of the emitter particle $\widetilde{ij}$ from the amplitude and attach it to the splitting matrix. Hence, the meaning of the indices depends on the type of splitting: if the emitter is a quark $\xi$ and $\xi'$ are Dirac indices, if the emitter is a gluon we deal with Lorentz indices.

Since we only need the helicity mixing terms, let us define an operator $\mathcal{R}$ that projects out the helicity mixing terms of any function that depends on the RP of particles $i$ and $j$:
\begin{equation}
	\label{eq:r-operator}
	\mathcal{R} f(\theta_i, \theta_j)
	\equiv f(\theta_i, \theta_j) - \sum_{\lambda_i, \lambda_j} f(\lambda_i, \lambda_j).
\end{equation}
Using this operator, we can define the splitting matrices in a general way:
\begin{equation}
	\label{eq:random-dipole-splitting-matrices}
	\widetilde{\mathbf{V}}_{ij,k}
	= C_{\ijToIJ} \, \mathcal{R} \left[ \widetilde{P}_{\ijToIJ} + \widetilde{S}_{\ijToIJ} \right]
\end{equation}
where $C$ is a color factor depending on the splitting, $C_{q \to qg} = C_F$, $C_{g \to gg} = C_A$ and $C_{g \to q \bar{q}} = T_R$. $\widetilde{P}$ and $\widetilde{S}$ are derived by looking at the soft and collinear limits of the real emission amplitude $\mathcal{A}_{n+1}$ in such a way that they are valid for both helicity eigenstates and RP.

If we first look at the collinear limit $p_i \parallel p_j$, we can rewrite the amplitude as follows:
\begin{equation}
	\lim_{p_i \parallel p_j} \mathcal{A}_{n+1} = g \mu^{\epsilon} \, \mathbf{T}_{\ijToIJ} \, \sum_{\lambda_{ij}} \text{Split}_{\ijToIJ}^{\lambda_{ij}} (\widetilde{p}_{ij}) \, \binom{u_{\lambda_{ij}} (\widetilde{p}_{ij})}{\epsilon_{\lambda_{ij}} (\widetilde{p}_{ij})}_{\xi} \, \mathcal{A}_n^\xi
\end{equation}
\pagebreak
The bracket indicates that depending on the emitter parton, we have to choose either a polarisation vector or a fermion spinor.
Since we do not have to worry about integrating the subtraction, we are much freer in the way we write our subtraction term: to this end, the $\text{Split}_{\ijToIJ}$ functions are defined using the usual Feynman rules for the splitting vertices (see table on the following page).
Furthermore, note that we rewrote the propagator of the emitter particle in terms of its polarisation sum,
\begin{equation}
	\frac{\sum_{\lambda_{ij}} \epsilon_{\lambda_{ij}}^* \epsilon_{\lambda_{ij}}}{2 p_i p_j},
\end{equation}
which is also the origin of the additional sum. Since this sum describes an internal degree of freedom, we do not replace it with an integral over a RP.
Squaring the amplitude yields
\begin{equation}
	\lim_{p_i \parallel p_j} \lvert \mathcal{A}_{n+1} \rvert^2
	= 4 \pi \alpha_s \mu^{2\epsilon} \, \bigl( \mathcal{A}_n^* \bigr)^\xi \, \mathbf{T}_{\ijToIJ}^2 \left[ \widetilde{P}_{\ijToIJ} \right]_{\xi \xi'} \, \mathcal{A}_n^{\xi'}
\end{equation}
where we defined
\begin{equation}
	\label{eq:collinear-function}
	\left[ \widetilde{P}_{\ijToIJ} \right]_{\binom{\alpha \beta}{\mu \nu}} (\widetilde{p}_{ij})
	= \sum_{\lambda, \lambda'} \binom{\bar{u}_\alpha^\lambda (\widetilde{p}_{ij})}{\epsilon_\mu^\lambda (\widetilde{p}_{ij})^*}
		{\text{Split}^\lambda}^* \text{Split}^{\lambda'}
		\binom{u_\beta^{\lambda'} (\widetilde{p}_{ij})}{\epsilon_\nu^{\lambda'} (\widetilde{p}_{ij})}.
\end{equation}
This term adequately describes the collinear part of our splitting matrix $\widetilde{\mathbf{V}}_{ij,k}$.

\begin{figure}
	\includegraphics[width=0.46\textwidth]{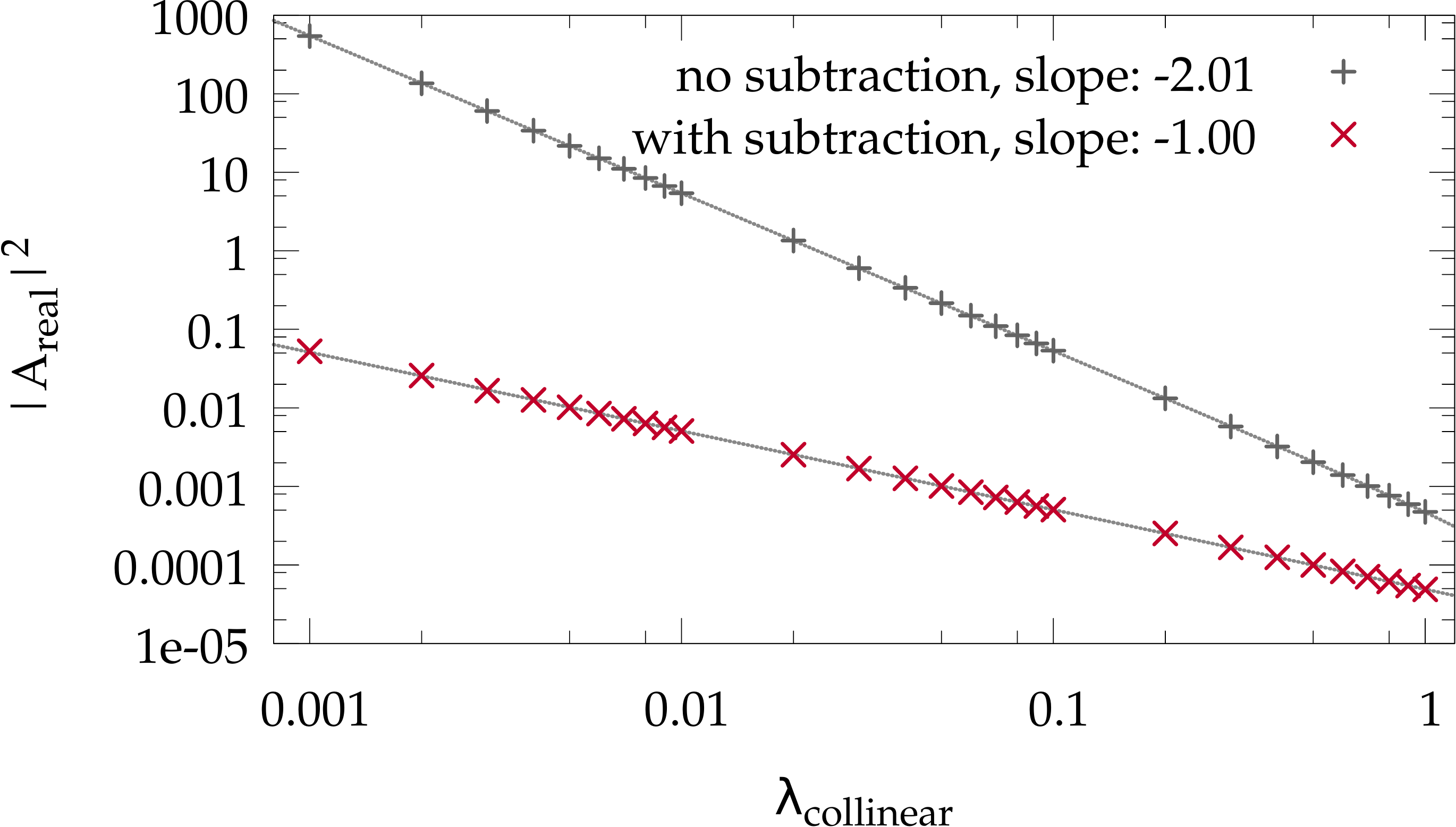} \hfill
	\includegraphics[width=0.46\textwidth]{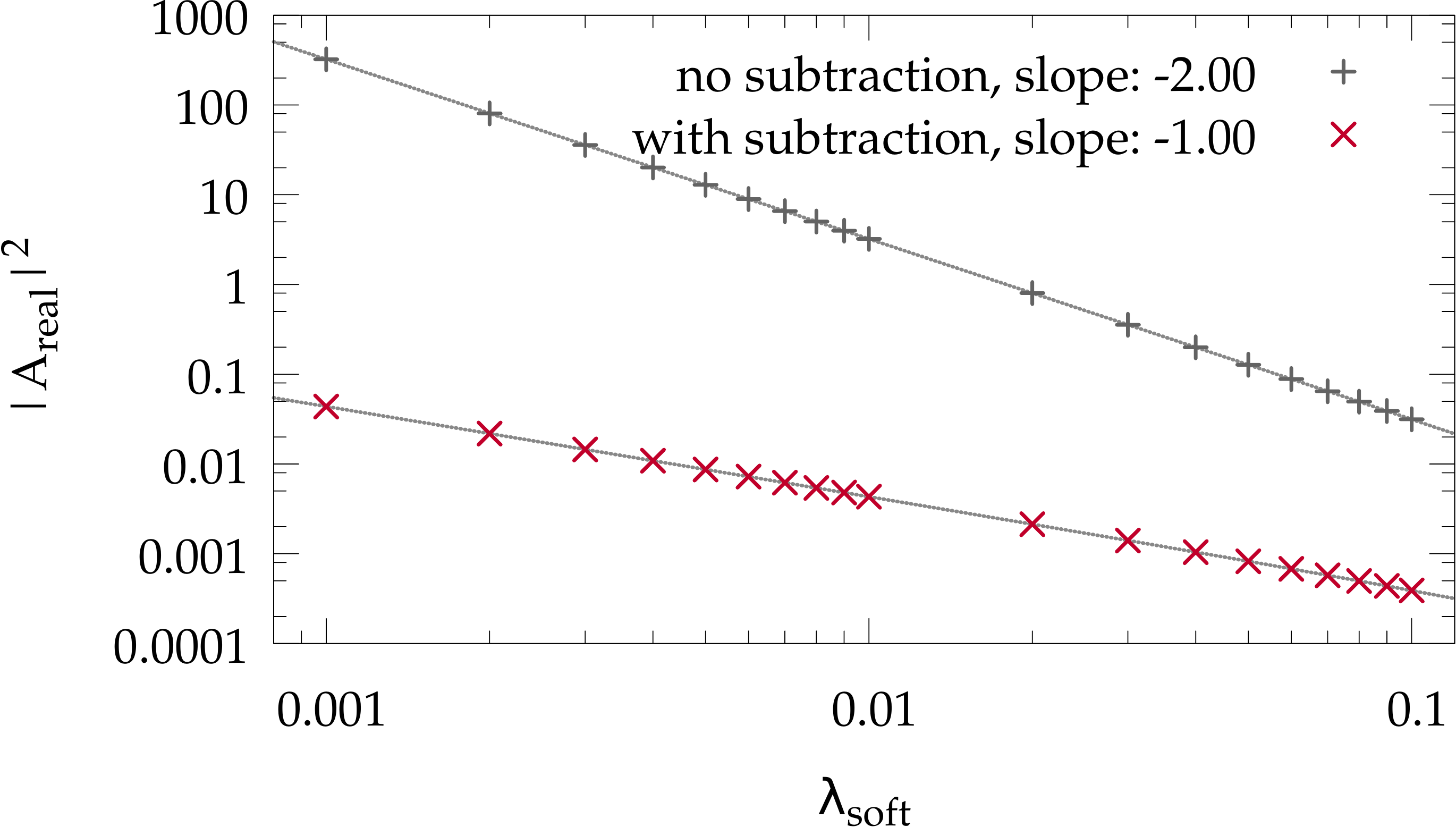}
	\caption{Subtraction in the collinear (left) and soft (right) limits. In both cases we subtract the singular $\frac{1}{\lambda^2}$ behaviour and are left with integrable $\frac{1}{\lambda}$ terms.}
	\label{fig:softcoll}
\end{figure}

Now we still need to treat the soft divergences. The soft expansion is well known:
\begin{equation}
	\label{eq:soft-limit}
	\lim_{p_j \to 0} \lvert \mathcal{A}_{n+1} \rvert^2
	=
	- 4 \pi \alpha_s \mu^{2\epsilon} \sum_{i=1}^n \sum_{\substack{ k = 1 \\ k \neq i }}^n \mathcal{A}_n^* \, \mathbf{T}_i \cdot \mathbf{T}_k \, S_{ij,k} (\epsilon_j) \, \mathcal{A}_n,
	\qquad
	\mathcal{A}_n = \binom{u (p_i)}{\epsilon (p_i)}_{\xi} \, \mathcal{A}_n^\xi
\end{equation}
where
\begin{equation}
\label{eq:soft-function}
	S_{ij,k} (\epsilon_j)
	= \frac{(p_i \cdot \epsilon_j^*) (p_i \cdot \epsilon_j)}{(p_i \cdot p_j)^2}
		- \frac{(p_i \cdot \epsilon_j^*) (p_k \cdot \epsilon_j) + (p_k \cdot \epsilon_j^*) (p_i \cdot \epsilon_j)}{(p_i \cdot p_j) (p_i \cdot p_j + p_j \cdot p_k)}
\end{equation}
In order to extract the soft term $\widetilde{S}_{\ijToIJ}$ for our splitting matrix, we have to beware of double counting since the collinear term $\widetilde{P}_{\ijToIJ}$ we defined earlier already contains some soft divergences. Hence, we investigate the collinear limit of the soft function $S_{ij,k}$ and the soft limit of the collinear function:
\begin{align}
	\text{collinear limit of } S_{ij,k}: &&
	\lim_{p_i \parallel p_j} S_{ij,k}
	&= \frac{(p_i \cdot \epsilon_j^*) (p_i \cdot \epsilon_j)}{(p_i \cdot p_j)^2}
	\\
	\text{soft limit of } \widetilde{P}_{\ijToIJ}: &&
	\lim_{p_j \to 0} \left[ \widetilde{P}_{\ijToIJ} \right]_{\binom{\alpha \beta}{\mu \nu}}
	&= \frac{(p_i \cdot \epsilon_j^*) (p_i \cdot \epsilon_j)}{(p_i \cdot p_j)^2}
		\binom{
			u_\alpha (p_i) \bar{u}_\beta (p_i)
		}{
			\epsilon_\mu (p_i)^* \epsilon_\nu (p_i)
		}.
\end{align}
Obviously, both limits agree%
\footnote{The soft function lacks the extra polarisations of the collinear function since they are hidden in $\mathcal{A}_n$, see eq.~\eqref{eq:soft-limit}.}
and we see that the collinear function already covers the first term of the soft function \eqref{eq:soft-function}.
Thus, we can cover all soft terms by defining the symbol $\widetilde{S}_{\ijToIJ}$ such that it only covers the second term of $S_{ij,k}$.\footnote{Note that we have to add the additional spinors/polarisations at this point to make the terms compatible.} The terms are shown in the following table:
\begin{center}
	\def\myfontsize{\small}
	\def\colora{black!4}
	\def\colorb{black!8}
	\fontsize{9pt}{11pt}\selectfont
	\renewcommand{\arraystretch}{1.3}
	\begin{tabular}{@{}cccc@{}}
		\rowcolor{\colora}
		& \myfontsize $\boldsymbol{q \to q g}$ & \myfontsize $\boldsymbol{g \to g g}$ & \myfontsize $\boldsymbol{g \to q \bar{q}}$
		\\
		\rowcolor{\colorb}
		\myfontsize $\boldsymbol{C_{\ijToIJ}}$ & $C_F$ & $C_A$ & $T_R$
		\\
		\rowcolor{\colora}
		&
			&
			$\dfrac{2}{2 p_i \cdot p_j} \Bigl[ \bigl( \epsilon_i \cdot \epsilon_j \bigr) \bigl( p_i \cdot (\epsilon_{ij}^\lambda)^* \bigr)$ &
			\\
				\rowcolor{\colora}
				&& $+ \bigl( \epsilon_j \cdot (\epsilon_{ij}^\lambda)^* \bigr) \bigl( p_j \cdot \epsilon_i \bigr)$ & \\
				\rowcolor{\colora}
				\multirow{-3}{*}{\myfontsize $\boldsymbol{\text{Split}_{\ijToIJ}^\lambda}$} &
				\multirow{-3}{*}{$\dfrac{1}{2 p_i \cdot p_j} \bar{u} (p_i) \gamma^\mu \epsilon_\mu (p_j) u^\lambda (\widetilde{p}_{ij})$} &
				$- \bigl( \epsilon_i \cdot (\epsilon_{ij}^\lambda)^* \bigr) \bigl( p_i \cdot \epsilon_j \bigr) \Bigr]$ &
				\multirow{-3}{*}{$\dfrac{1}{2 p_i \cdot p_j} \bar{u} (p_i) \gamma^\mu \epsilon_\mu^\lambda (\widetilde{p}_{ij})^* v (p_j)$}
		\\
		\rowcolor{\colorb}
		&
			$- \frac{(p_i \cdot \epsilon_j^*) (p_k \cdot \epsilon_j) + (p_k \cdot \epsilon_j^*) (p_i \cdot \epsilon_j)}{(p_i \cdot p_j) (p_i \cdot p_j + p_j \cdot p_k)}$ &
			$- \frac{(p_i \cdot \epsilon_j^*) (p_k \cdot \epsilon_j) + (p_k \cdot \epsilon_j^*) (p_i \cdot \epsilon_j)}{(p_i \cdot p_j) (p_i \cdot p_j + p_j \cdot p_k)}$ &
		\\
		\rowcolor{\colorb}
		\multirow{-2}{*}{\myfontsize $\boldsymbol{\Bigl[ \widetilde{S}_{\ijToIJ} \Bigr]_{\xi \xi'}}$}
		&
			$\times \ u_\alpha (p_i) \bar{u}_\beta (p_i)$ &
			$\times \ \epsilon_\mu (p_i)^* \epsilon_\nu (p_i) - (i \leftrightarrow j)$ &
			\multirow{-2}{*}{$0$}
	\end{tabular}
\end{center}

We have verified that $\widetilde{P}$ and $\widetilde{S}$ correctly subtract the collinear and soft poles in the respective limits when using random polarisations.
Fig.~\ref{fig:softcoll} shows two example plots for subtracted and unsubtracted real emission matrix elements in both limits.
The parameter $\lambda_{\text{collinear}}$ is proportional to the invariant $s_{ij}$ of the collinear particle pair, and $\lambda_{\text{soft}}$ is a scaling parameter for the soft momentum.

\section{Conclusions}

The previous section showed how to derive an additional subtraction term that takes care of local poles produced by the new helicity mixing terms that are inherent to random polarisations which provide a great speed-up for numerical calculations of jet cross sections.
The above table in combination with equations \eqref{eq:random-dipole-subtraction-term} -- \eqref{eq:random-dipole-splitting-matrices} and \eqref{eq:collinear-function} shows all the formulas necessary to extend the existing final-final dipoles to RP.
In our publication \cite{Goetz:2012uz} we also derived the other three cases by using crossing symmetry.

Furthermore, we only discussed the massless case in this talk.
The general massive case is described in our publication and requires a modification of the parametrisations for $\widetilde{p}_{ij}$ and $\widetilde{p}_k$ for the quasi-collinear limit, similar to \cite{Catani:2002hc}.

Finally, let us stress one more important fact. Although our method is based on the dipole formalism, we never require any knowledge of the original subtraction term $\drm \sigma^{\text{A}}$ which means that our extension can be used with any subtraction scheme as long as it is based on helicity summed squared amplitudes.

\bibliography{RADCOR-proceedings-goetz}
\bibliographystyle{h-physrev}

\end{document}